\documentclass[a4paper]{article}

\usepackage{INTERSPEECH2022}

\usepackage{float}
\usepackage{tikz}
\usepackage{collcell}
\usepackage{diagbox}
\usepackage{rotating}
\usepackage{amsmath}

\graphicspath{ {./images/} }

\title{Relative Acoustic Features for Distance Estimation in Smart-Homes}
\name{Francesco Nespoli$^1$$^2$, Daniel Barreda$^1$ and Patrick A. Naylor$^2$}
\address{
  $^1$Nuance Communications UK\\
  $^2$Imperial College London}
\email{francesco.nespoli@nuance.com, daniel.barreda@nuance.com, p.naylor@imperial.ac.uk }

\begin{document}

\maketitle
\begin{abstract}
  Any  audio  recording  encapsulates  the  unique  fingerprint  of the  associated  acoustic  environment,  namely  the  background noise and reverberation.  Considering the scenario of a room equipped with a fixed smart speaker device with one or more microphones and a wearable smart device (watch, glasses or smartphone),  we  employed  the  improved  proportionate normalized least mean square adaptive filter to estimate the relative room impulse response mapping the audio recordings of the two devices. We performed  inter-device distance estimation by exploiting a new set of features obtained extending the definition of some acoustic attributes of the room impulse response to its relative version. In combination with the sparseness measure of the estimated relative room impulse response, the relative features allow precise inter-device distance estimation which can be exploited for tasks such as best microphone selection or acoustic scene analysis. Experimental  results from simulated rooms of different dimensions and reverberation times demonstrate  the  effectiveness of  this computationally  lightweight  approach  for  smart  home acoustic ranging applications.
\end{abstract}

% INTRODUCTION
\section{Introduction}
    The task of indoor localization with wearable devices is a very challenging engineering problem which has gained more attention in recent years. Commonly, localization techniques employing smart devices are composed of two blocks: the first estimates geographic features such as the relative distance and angles between devices; the second aggregates these measurements to produce the putative location of the target device \cite{Liu}. In this paper we focus on the first block particularly on the task of distance estimation in indoor environments, also known as indoor-ranging. In the literature, this problem has been tackled with many different technologies such as WiFi \cite{ijgi9110627}, radio frequency \cite{Porcino2002}, light beams \cite{light}, inertial sensors \cite{inertial} and computer vision \cite{app12031354}. However, all of the aforementioned methods have significant drawbacks. Radio frequency, inertial sensors and light beams require ad-hoc infrastructure and expensive sensors. Computer vision requires huge image data sets, indoor light conditions severely affect the performance \cite{lighttool}, it might incur privacy issues \cite{privacy} and cannot be applied in the context of smart devices due to its high computational cost \cite{computational}. Because of these reasons, it is relevant to focus on acoustic localization which achieves relatively high accuracy and low latency, requiring only microphones and speakers usually embedded in modern smart devices. A popular approach for indoor-ranging in the acoustic domain consists of measuring the time of flight (ToF), which is the time a signal takes to travel from transmitter to receiver. In this way, knowing the emission time of the transmitter, it is possible to estimate the inter-device distance. However, this technique does not achieve very good results mainly due to poor ToF estimation \cite{Liu}. In the context of smart-homes, many devices are interconnected and can be remotely controlled. Commonly, a smart speaker equipped with microphones is used to control the other devices in the house. Moreover, most people wear or carry one or more smart devices equipped with a microphone such as a watch or glasses. 
    In this scenario, we present a novel acoustic ranging system based on a set of acoustic features extracted from the relative room impulse response (\(\overleftrightarrow{RIR}\)) estimated between the speaker and the wearable device in the time domain. This feature set has been inspired by acoustic features commonly extracted from the room impulse response (\(RIR\)) but modified to be relative measures. Specifically, the clarity index (\(C_{50}\)) and the direct to reverberant ratio (\(DRR\)), which take into account both source-receiver distance and room acoustics, and the reverberation time (\(T_{60}\)) have been used. In combination with the signal to reverberation ratio (\(SRR\)) and a sparseness (\(S\)) measure of the \(\overleftrightarrow{RIR}\), we employed the optimized distributed extreme gradient boosting algorithm (XGBoost) \cite{xgboost} with regression trees to estimate the distance between the smart speaker and wearable device, and compared its performance with a ToF baseline employing the Generalized Cross Correlation (GCC) with phase transform (PHAT) \cite{GCC} and with a dense neural network (DNN) architecture. Finally, we demonstrate the usefulness and importance of the relative features for the particular regression model predictions using SHapley Additive exPlanations (SHAP)~\cite{shap}.
% END INTRODUCTION

\section{Problem Formulation}
In this paper, the mathematical notation used is the following:
    \raggedbottom
    \begin{gather*} 
        \textbf{x}(n) = [ x(n), x(n-1), ... , x(n-L+1) ]^T~\,, \\
        \textbf{y}(n) = [ y(n), y(n-1), ... , y(n-L+1) ]^T~\,, \\
        \boldsymbol\nu(n) = [ \nu(n), \nu(n-1), ... , \nu(n-L+1) ]^T~\,, \\
        \textbf{h} = [ h_0, ... , h_{L-1} ]^T~\,, \\
        \textbf{\^{h}}(n) = [ \hat{h}_0(n), ... , \hat{h}_{L-1}(n) ]^T~\,, \\
        {y}(n) = \textbf{h}^T\textbf{x}(n)+\nu(n)~\,,
    \end{gather*}
    where  \(x(n)\), \(y(n)\) and \(\nu(n)\) are audio samples from a smart speaker (near-field), a wearable device (far-field) microphones and a noise component, \textbf{h} and \textbf{\^{h}} are the time-invariant ground-truth and the estimated \(\overleftrightarrow{RIR}\), \emph{L} is the length of the ground truth \(\overleftrightarrow{RIR}\), \(n\) is the time index and \([\cdot]^{T}\) is the transpose operator. 

\subsection{Relative acoustic features extraction}
    In this section, we explain and motivate the use of \(C_{50}\), \(T_{60}\), \(DRR\), \(SRR\) and \(S\) for the distance estimation problem. Conventionally, \(C_{50}\), \(T_{60}\) and \(DRR\) are estimated from \(RIR\)s. However, in most smart home applications \(RIR\) measurements are not available. Given that acoustic features such as \(C_{50}\) and \(DRR\)  are directly related to source-receiver distance, conventional acoustic ranging systems need blind estimates obtained from reverberant signals \cite{pablo}. However, blind approaches may be inaccurate leading to poor distance estimations. Here, we introduce the relative acoustic features (\(\overleftrightarrow{\cdot}\)) which are calculated from the \(\overleftrightarrow{RIR}\) estimate \textbf{\^{h}} as follows: 

    \begin{equation}
        \overleftrightarrow{C_{50}} = 10\cdot{\log _{10}} \Bigg( \frac{\sum_{0}^{n_{50}}\textbf{\^{h}}^2(n) }{\sum_{n_{50}}^{\infty}\textbf{\^{h}}^2(n) } \Bigg)~\,,
    \end{equation}
    
    \begin{equation}
        \overleftrightarrow{DRR} =  10\cdot{\log _{10}} \Bigg( \frac{\sum_{n_d-n_0}^{n_d+n_0}\textbf{\^{h}}^2(n) }{\sum_{n_d+n_0}^{L-1}\textbf{\^{h}}^2(n)} \Bigg)~\,,
    \end{equation}
    where \(n_d\) and \(n_0\) (2) are the sample index of the peak of the direct sound arrival and the number of samples corresponding to a small temporal window (2.5 ms in our case) and \(n_{50}\) (1) is the sample index for 50~ms. The \(\overleftrightarrow{T_{60}}\), defined as the time required by the \(\overleftrightarrow{RIR}\) energy to decay from $-5$~dB to $-60$~dB,  was calculated from the relative energy decay curve 
    
    \begin{equation}
        \overleftrightarrow{E}(n) = 10\cdot{\log _{10}}\left({{{\int_n^\infty  \textbf{\^{h}}^2(\xi){\rm{d}}\xi } \over {\int_0^\infty \textbf{\^{h}}^2(\xi){\rm{d}}\xi }}} \right)~\,,
    \end{equation}
    by performing linear interpolation
    
    \begin{equation}
       \overleftrightarrow{T_{60}} = \frac{-60~\textrm{dB}}{A}~\,,
    \end{equation}
    where \(A\) is the slope coefficient from the interpolated line (in dB/s). We also exploited the sparseness coefficient of the estimated filter \textbf{\^{h}} \cite{Loganathan2008ASC}, defined as
    \begin{equation*}
        \overleftrightarrow{S}(n) =  \frac{L}{L- \sqrt{L}} \Bigg( 1 - \frac{||\textbf{\^{h}}(n)||_1}{\sqrt{L}||\textbf{\^{h}}(n)||_2} \Bigg)~\,,
    \end{equation*}
    where \(||\cdot||_1\) and \(||\cdot||_2\) define the \(l_1\)-norm and \(l_2\)-norm respectively, since \(\overleftrightarrow{S}(n)\) depends on room acoustics and source-receiver distance. Finally, \(SRR\) was calculated following \cite{segSRR} as
    \begin{equation}
        SRR =  10\cdot{\log _{10}} \Bigg(\frac{||\textbf{x}(n)||_{2}^{2}}{||\textbf{y}(n)||_{2}^{2}} \Bigg)~\,.
    \end{equation}

\subsection{Improved proportionate normalized least mean square algorithm overview}
    In this section we briefly review the improved  proportionate normalized least mean square (IPNLMS) adaptive filter firstly introduced in \cite{Benesty2002} and now used in \(\overleftrightarrow{RIR}\) estimation for the smart home scenario. The role of IPNLMS is to produce the \(\overleftrightarrow{RIR}\) estimate \(\textbf{\^{h}}(n)\) which describes the relationship between the smart speaker and the wearable device recorded signals. The adaptive filter iteratively minimizes the error signal
    \begin{equation}
    e(n) = y(n) - \textbf{\^{h}}^T(n-1)\textbf{x}(n)
    \end{equation}
    using
    \begin{equation}
     \textbf{\^{h}}(n) = \textbf{\^{h}}(n-1) + \mu \frac{ e(n) \textbf{K}(n-1) \textbf{x}(n)}{\textbf{x}(n)^T \textbf{K}(n-1) \textbf{x}(n) +\delta}~\,,
    \end{equation}
    where $\mu$ is the adaptation step and $\delta$ is the regularization factor. The role of \textbf{K} is to update the filter taps proportionally to their relative magnitude and
    \begin{equation}
     \textbf{K}(n-1) = \textrm{diag}(k_0(n-1), ... , k_{L-1}(n-1))~\,,\\
    \end{equation}
    \begin{equation}
     k_l(n) =  \frac{1 - \alpha}{2L} + (1+\alpha) \frac{|\hat{h_l}(n)|}{2||\textbf{\^{h}}(n)||_1+\epsilon}, l=0,...,L-1\,.
    \end{equation}
   Here $\alpha$ is a real number in the interval \([-1,+1]\) and $\epsilon$ is a small number necessary to avoid division by zero.

\subsection{Distance regression and model explainability}
    Distance estimation has been carried out with the extreme gradient boosting algorithm XGBoost with decision tree regressor \cite{xgboost} as the base model. We chose this model because it has been demonstrated to perform better on tabular data inputs compared with neural networks \cite{tabular} and because of its explainability with respect to the input features. We shaped the distance estimation problem as a supervised task by providing, at training time, the ground truth values for speaker-wearable device distance. We trained the model on standardized input features employing the mean squared error (MSE) loss for a number of epochs established at run time with the early stopping criterion of 5 epochs without improvements on the validation loss. The optimal model hyperparameters has been chosen using a grid search approach. We compared XGBoost with a DNN with 5 layers trained with the MSE loss for a number of epochs determined with the same early stopping condition mentioned above. For comparison, we also provide a baseline estimation obtained by computing the ToF with GCC-PHAT.  Finally, we employed SHAP \cite{shap} to establish the importance of the input features on the model prediction for both XGBoost and DNN.

\section{Experiments and Results}

 Results are presented for simulated data in three rooms:
    
    \begin{itemize}
      \item Small: [5, 3.5, 2.5]~m, \(T_{60}\)~=~0.5~s
      \item Medium: [7, 4.5, 2.5]~m, \(T_{60}\)~=~0.7~s
      \item Large: [8, 6.5, 2.5]~m,  \(T_{60}\)~=~0.9~s
    \end{itemize}
    where we used broad-band white noise and speech input signals. 

% PROBLEM FORMULATION
\subsection{Description of the experiments}
    In order to investigate the performance of distance estimation exploiting relative features we simulated reverberant rooms with the image source model~\cite{pyroom}. For each simulated room, we positioned a wearable device in 30000 random positions and the position of the smart speaker was fixed. For each position of the wearable device, we simulated the reverberant sound signal recorded by both the smart speaker and the wearable device and we performed \(\overleftrightarrow{RIR}\) estimation using IPNLMS as described in Section 2.2. For the simulations, we employed two types of sound signals. First, we used broad-band white noise for optimal full-band spectral excitation. Second, we tested our model with speech signals of different duration randomly selected from the LJSpeech data set \cite{ljspeech17}. Both simulated conditions are closely related to real-life scenarios for different operational modes of the smart speaker. For example, when the speaker is in silent mode, meaning it is not executing any command or playing any sound, an inaudible noise would be played in order to acoustically locate the wearable device. Note that in this case spectrally shaped noise in the frequency band 15-20 kHz would be appropriate to not disturb human listeners or interfere with speaker and microphone operations \cite{Dokmani}.  With this band-filtered signal it is possible to retain full-band excitation by resampling the audio signals before \(\overleftrightarrow{RIR}\) estimation \cite{vaidyanathan1993multirate}. On the other hand, in the operational mode when the device is acoustically interacting with the environment, typically in a dialogue, acoustic speech signals would be used for the ranging task. 
    
    \subsection{Acoustic channel estimation}
         Figure 1 shows the estimation convergence and results for a relative room impulse response estimation with the speaker operating in silent mode.
        \vspace*{-\baselineskip}
        \begin{figure}[H]
        \centerline{\includegraphics[scale=0.22]{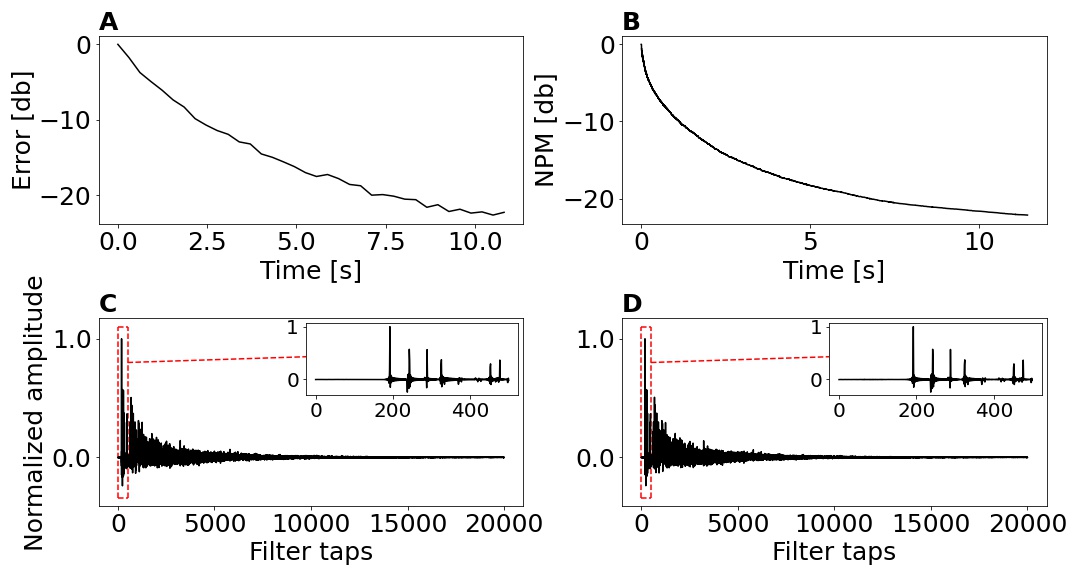}}
        \caption{Acoustic channel estimation. \textbf{A}: Estimation error, \textbf{B}: Normalized Projection Misalignment (NPM), \textbf{C}: Ground truth relative RIR, \textbf{D}: Estimated relative RIR.}
        \label{fig1}
        \end{figure}
        The ground truth \(\overleftrightarrow{RIR}\) shown in Fig.~1C was calculated as
            \begin{equation}
            \overleftrightarrow{RIR} =   \mathcal{F}^{-1}  \Bigg( \frac{RIR_{far}}{RIR_{close}} \Bigg)~\,,
            \end{equation}
        where \(RIR_{close},RIR_{far}\), are the room impulse responses from the source to the smart speaker and wearable device microphones respectively, calculated for an inter-device distance of 1.55~m in the frequency domain. Figure 1A shows the evolution of the error across iterations for an IPNLMS filter with  $\alpha=0$,  $\mu=0.3$ for the first 4.5~s and $\mu=0.1$ subsequently, white noise input with SNR~=~30~dB and a ground truth \(\overleftrightarrow{RIR}\) length of \(L\)~=~20000. The similarity between ground truth and estimated filter has been assessed by computing the normalized projection misalignment \cite{npm, npm2} as shown in Fig.~1B.

    \subsection{Distance estimation for white noise inputs}
        After extracting the \(\overleftrightarrow{RIR}\) and computing the features introduced in Section 2.1, the correlation matrix was calculated among all the relative features and the inter-device distance.
        % \documentclass[12pt]{article}
% \usepackage{tikz}
% \usepackage{collcell}
% \usepackage{diagbox}
% \usepackage{rotating}
% \usepackage{amsmath}
% % \usepackage{INTERSPEECH2021}
% \usepackage{float}

        %The min, mid and max values
        \newcommand*{\MinNumber}{-0.79}%
        \newcommand*{\MidNumber}{0.9} %
        \newcommand*{\MaxNumber}{1.00}%
        
        %Apply the gradient macro
        \newcommand{\ApplyGradient}[1]{%
                \ifdim #1 pt > \MidNumber pt
                    \pgfmathsetmacro{\PercentColor}{max(min(100.0*(#1 - \MidNumber)/(\MaxNumber-\MidNumber),100.0),0.00)} %
                    \hspace{-0.33em}\colorbox{green!\PercentColor!yellow}{#1}
                \else
                    \pgfmathsetmacro{\PercentColor}{max(min(100.0*(\MidNumber - #1)/(\MidNumber-\MinNumber),100.0),0.00)} %
                    \hspace{-0.33em}\colorbox{red!\PercentColor!yellow}{#1}
                \fi
        }
% \vspace*{-\baselineskip}
\begin{table}[H]

   \newcolumntype{R}{>{\collectcell\ApplyGradient}c<{\endcollectcell}}
    \renewcommand{\arraystretch}{0}
    \setlength{\fboxsep}{1.25mm} % box size
    \setlength{\tabcolsep}{0pt}

\begin{tabular}{c*{7}{R}}

        \multicolumn{1}{c}{} &  \multicolumn{1}{c}{$\overleftrightarrow{T_{60}}$} &
        \multicolumn{1}{c}{$\overleftrightarrow{C_{50}}$} &  \multicolumn{1}{c}{$\overleftrightarrow{DRR}$} &
        \multicolumn{1}{c}{$SRR$} & 
        \multicolumn{1}{c}{$\overleftrightarrow{S}$} & 
        \multicolumn{1}{c}{distance} \\[0.4em]
 $\overleftrightarrow{T_{60}}$ & +1.00& - 0.32& - 0.43& +0.18& - 0.50& +0.49\\
 $\overleftrightarrow{C_{50}}$ & - 0.32& +1.00& +0.83& - 0.27& +0.91& - 0.56\\
 $\overleftrightarrow{DRR}$    & - 0.43& +0.83& +1.00& - 0.29& +0.88& - 0.79\\
 $SRR$                         & +0.18& - 0.27& - 0.29& +1.00& - 0.32& +0.29\\
 $\overleftrightarrow{S}$      & - 0.50& +0.91& +0.88& - 0.32& +1.00& - 0.78\\
 distance \( \)                & +0.49& - 0.56& - 0.79& +0.29& - 0.78& +1.00\\
\end{tabular}
\caption{Cross correlation matrix for white noise input signal}
\end{table}
\vspace*{-\baselineskip}
        Table 1 displays strong linear correlations for all the extracted relative features with the speaker-wearable distance particularly for \(\overleftrightarrow{DRR}\) and the sparseness coefficient suggesting that this set of features embodies the inter-device distance information. Speaker-wearable device distance estimation has been carried out employing the gradient boosting decision trees model extensively presented in Section 2.3. We used 80\% of the available data as the training set, 10\% for validation and the remainder for testing. After training, we computed the relative error metric (\(RE\)) on the test set using
        \begin{equation}
         RE[\%] = \frac{ \sqrt{(d_{tar}-d_{pred})^2}}
         {d_{tar}}\times100~\,,
        \end{equation}
        where \(d_{tar}, d_{pred}\) are the ground truth and the predicted distances respectively. We obtained an average \(RE\) of 9.01\% meaning that on average the error on the estimated distance is approximately 9\% of the actual inter-device distance. We compared our method with a ToF algorithm based on GCC-PHAT and a DNN with the same input features as XGBoost. To calculate the GCC-PHAT estimation, we divided the wearable device and smart speaker audio signals into 300 ms time windows and computed the cross correlation across all frames. We then extracted the time index corresponding to the maximum value of the correlation for each time window and estimated the distance by multiplying it with the speed of sound extracted from the room simulation. By averaging the estimated distances across all frames, we calculated the GCC-PHAT distance estimate. We obtained an \(RE\) of 11.7\%, more than 25\% higher compared to the proposed method. Figure 2 shows further investigations of our method by comparing the performance ofhttps://www.overleaf.com/project/62a9fba3399eb92950f5462a XGBoost in rooms of different dimensions and reverberation times with GCC-PHAT and the DNN.
        \vspace*{-\baselineskip}
        \begin{figure}[H]
        \centerline{\includegraphics[scale=0.4]{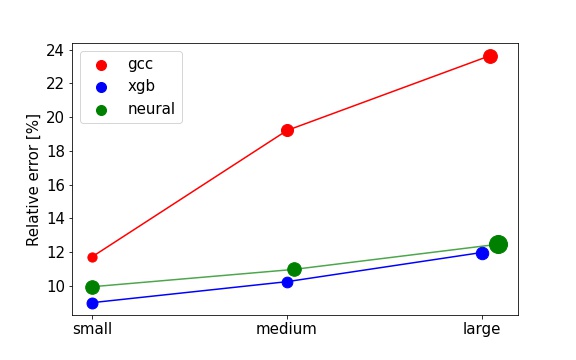}}
        \centering
        \begin{tabular}{ |p{1.75cm}|p{1.5cm}|p{1.5cm}|p{1.5cm}|  }
        \hline
        \multicolumn{4}{|c|}{Relative error [\%] : mean(variance)} \\
        \hline
        Algorithm&Small&Medium&Large \\
        \hline
        XGBooost & 9.01(0.24) & 10.25(0.23) & 11.98(0.32)\\
        Neural Net & 9.96(0.37)   & 10.98(0.38) & 12.47(0.66)\\
        GCC-PHAT  & 11.7(0.18)   & 19.2(0.29) & 23.7(0.38)\\
        \hline
        \end{tabular}
        \caption{Relative error comparison for GCC-PHAT (gcc), XGBoost (xgb) and a DNN (neural) for white broad-band inputs.}
        \label{fig3}
        \end{figure}
        The value of the relative error for each room has been obtained by averaging the results from 10 Monte Carlo simulations. The diameter of the dots, represents the variance across simulations. Finally, we show the relative importance of the input features for the dense network and XGBoost algorithms. Figure 3 shows the comparison of the SHAP values for the two algorithms.   Interestingly, both the DNN and the random forest display similar ordering of the features  importance showing that \(\overleftrightarrow{S}\) and \(\overleftrightarrow{T_{60}}\) play the most important role for inter-device distance estimation.
        \begin{figure}[!ht]
        \centerline{\includegraphics[scale=0.62]{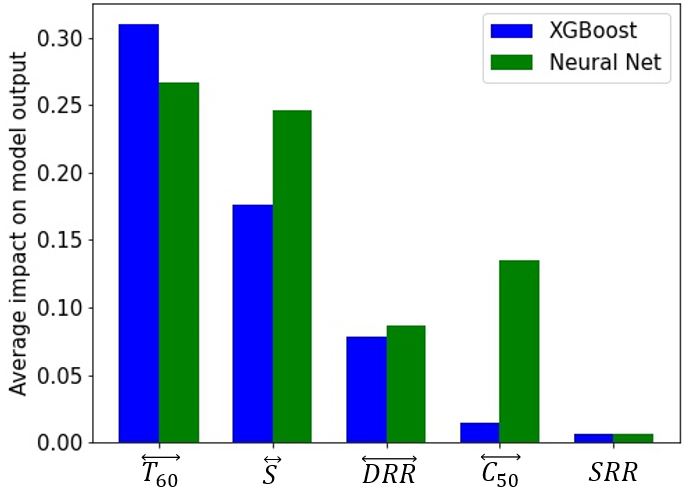}}
         \caption{SHAP values for XGBoost and the DNN calculated for the small room and white noise inputs.}
        \label{fig4}
        \end{figure}
However, \(\overleftrightarrow{C_{50}}\) and \(SRR\) appear not to be useful parameters for distance estimation. To test this hypothesis, we retrained new models excluding \(SRR\) from the input features and compared the distributions of the relative error with the models using \(SRR\). Student t-tests for XGBoost predictions evaluated the relative error distributions not statistically different for all rooms (small: p-value~=~0.34, medium: p-value~=~0.31, large: p-value~=~0.68. Similar results were obtained for the DNN). Finally, testing the error distributions obtained with and without the \(\overleftrightarrow{C_{50}}\), we obtained statistical difference for all rooms and both of the algorithms, suggesting \(\overleftrightarrow{C_{50}}\) was a useful feature for distance estimation (Student t-tests for XGBoost predictions per-room. Small: p-value~=~0.015, medium: p-value~=~0.001, large: p-value~\(<\)~0.001. Similar results for the DNN). 
    
\subsection{Distance estimation with speech signals}
    In this section we present distance estimation results obtained using speech input signals from the LJSpeech corpus \cite{ljspeech17}. 
        \vspace*{-\baselineskip}
        \begin{figure}[!ht]
        \centerline{\includegraphics[scale=0.4]{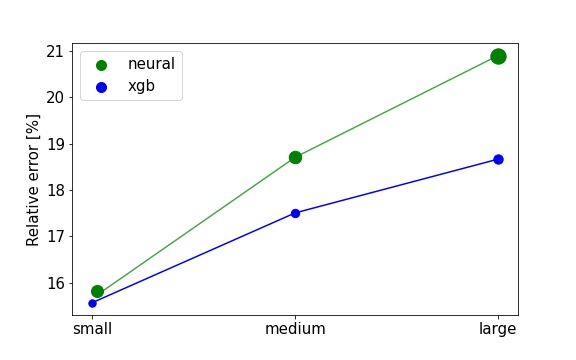}}
        \centering
        \begin{tabular}{ |p{1.75cm}|p{1.5cm}|p{1.5cm}|p{1.5cm}|  }
        \hline
        \multicolumn{4}{|c|}{Relative error [\%] : mean(variance)} \\
        \hline
        Algorithm&Small&Medium&Large \\
        \hline
        XGBooost & 15.57(0.22) & 17.51(0.28) & 18.67(0.35)\\
        Neural Net & 15.82(0.58)   & 18.71(0.62) & 20.90(0.94)\\
        
        \hline
        \end{tabular}
        \caption{Relative error comparison for XGBoost (xgb) and a DNN (neural) for speech input signals.}
        \label{fig5}
        \end{figure}
         We used speech utterances with minimum length of 10~s which have been obtained by processing the original audios with the WebRTC voice activity detector (VAD) from Google. This VAD choice was motivated due to the computationally inexpensiveness of the algorithm therefore being optimal in on-board applications for smart devices with limited computational power. For each utterance, speech segments of 30~ms were extracted and then aggregated into a single audio. Finally, if the new audio signal was longer than 10~s, it was used for room acoustic simulations. Results for the same three rooms as in the case of white broad-band noisy inputs are presented in Fig.~4. The cross-correlation baseline has not been included because it scored \(RE\)s higher than 100\%. 
        \begin{figure}[!ht]
        \centerline{\includegraphics[scale=0.62]{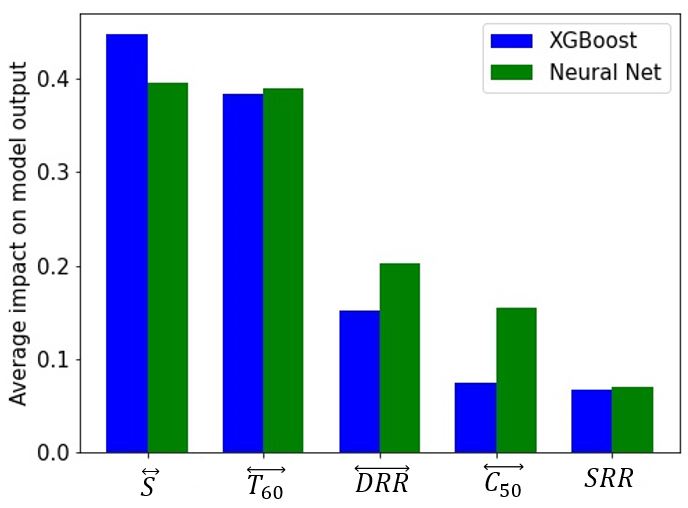}}
         \caption{SHAP values for XGBoost and the DNN calculated for the small room and speech inputs.}
        \label{fig6}
        \end{figure}
        Finally, in Fig.~5 we present the SHAP values for the case of speech inputs. Interestingly, for speech utterances all extracted features were statistically useful for distance estimation. Student t-tests for XGBoost and DNN predictions generated p-values \(<\)~0.05 for all rooms and subsets of features including $\overleftrightarrow{C_{50}}$, $\overleftrightarrow{DRR}$ and $SRR$.

\section{Conclusions}
   We have first introduced relative acoustic characteristic parameters and presented a novel ranging system based on relative acoustic features extracted from the estimated \(\overleftrightarrow{RIR}\) with the IPNLMS adaptive filter. The proposed method does not require the \(RIR\) nor any model to blindly estimate the acoustic properties of the room. We demonstrated that by extracting acoustic features directly from the \(\overleftrightarrow{RIR}\), it is possible to estimate accurately the distance between a smart speaker and a wearable device in acoustically different indoor environments. Our method can be adopted in smart home environments with both broad-band white noise and speech input signals for applications such as distributed microphone array geometry reconstruction and acoustic scene analysis. Further steps include a second neural network trained directly on the \(\overleftrightarrow{RIR}\) which, in combination with constraints on the current distance estimation taking into account previous predictions, pave the way for wearable device acoustic tracking applications.
    
\section{Acknowledgements}
This project was funded by the European Union’s Horizon 2020 program under the Marie Skłodowska-Curie grant agreement No 956369.

% Force references to new page
\clearpage

\bibliographystyle{IEEEtran}

\bibliography{sapstrings,new_bib}

%\begin{thebibliography}{9}
%\end{thebibliography}

\end{document}